\documentstyle[12pt,subeqn]{article}
\def\displayfrac#1#2{\frac{\displaystyle #1}{\displaystyle #2}}
\begin{document}
\baselineskip=2pc

\title{PCAC in Nuclear Medium and the Lovelace - Shapiro - Veneziano formula}

\author{J.Pasupathy\\Centre for Theoretical Studies\\
Indian Institute of Science\\Bangalore 560012}

\date{ }
\maketitle

\begin{abstract}
A simple way to enforce the Adler zero condition for pion
amplitudes in the nuclear medium is to use the Lovelace
quantization condition but with modified Regge Slope $\alpha^{\prime}$
This  latter is related to change in the gluon condensate.
Increasing nuclear density leads to a relative increase in the
Regge Slope $\alpha^{\prime}$  Denoting this increase by a scale factor
$\lambda$, the drop in the $\rho$-mass, $\Delta-N$ mass
difference, increase in $\pi\pi$ scattering length, decrease of
pion decay constant etc. are simply related to $\lambda$.
\end{abstract}

\vspace{0.3cm}
PACS INDEX:  11.40 Ha, 11.30 Rd, 12.40 Nn, 21.65. +f

\vspace{0.3cm}

\newpage
\noindent
There is substantial evidence that nucleon properties are
modified inside a nucleus.  These are commonly the quenching of
the magnetic moments and Gamow-Teller matrix elements,
Nolen-Schiffer anomaly, nuclear EMC effect etc.  The question of
plausible pion and kaon condensation, restoration of chiral
symmetry are also of great interest.  Further recent experiments
on heavy ion collisions  in which large dilepton cross-sections
are observed, have been interpreted as due to ``dropping'' of
vector meson masses in the nuclear medium\cite{1}.

Here we shall address some of these questions using the
following two principles.

1) Even if chiral symmetry is partially restored in the nuclear
medium, as long as it remains a spontaneously broken symmetry
the pion retains its Nambu-Goldstone character.  PCAC then
demands that an amplitude with a pion on the external line
must vanish when the external pion's four momentum goes to zero
(Adler Zero). 

2)  Hadron masses are determined by the slope $\alpha^{\prime}$ of Regge
trajectories or the string tension which can be related to the
distribution of gluonic fields in the physical vacuum state.
The gluonic fields are modified in the nuclear medium and lead
to a reduction in the string tension or an effective increase in
the value of $\alpha^{\prime}$.

These two principles find an easy expression in the
Lovelace-Shapiro-Veneziano\cite{2,3} formula.  Let us briefly
recapitulate some of the old analysis, using currently known
experimental information.  The amplitude\break
$\pi^+(p_1)+\pi^-(p_2)\rightarrow \pi^+(p_3)+\pi^-(p_4)$ can be
represented by a single term formula\cite{2,3}.
     \begin{eqnarray}
     A(s,t)=-\beta\displayfrac{\Gamma(1-\alpha(s))\Gamma(1-\alpha(t))}
     {\Gamma(1-\alpha(s)-\alpha(t))}
     \end{eqnarray}

\noindent
Here $\alpha(s)$ is the linear Regge trajectory represented by 
    \begin{eqnarray}
    \alpha(s) = \alpha_0+\alpha^{\prime}s
    \end{eqnarray}

\noindent
and the Mandelstam invariants are
     \begin{eqnarray}
     s=(p_1+p_2)^2\quad\quad t=(p_1-p_3)^2\quad\quad u=(p_1-p_4)^2
     \end{eqnarray}

\noindent
The constant $\beta$ will be determined below.  As pointed out by
Lovelace\cite{2}, setting $p_{1\mu}=0$ and demanding that
$A(s,t)$ vanish, leads to the quantization condition
    \begin{eqnarray}
    \alpha\left(m^2_{\pi}\right) =1/2
    \end{eqnarray}

\noindent
The intercept $\alpha_0$ and the slope $\alpha^{\prime}$ in
eqn.(2) can be determined using the spin 1, $\rho(770)$ and
the spin 3, $\rho_3(1690)$ masses.  We have \cite{4}.
     \begin{eqnarray}
     m_{\rho}=768.5\pm 0.6 MeV\;\;\; m_{\rho_{3}}=1691\pm 5 MeV
     \end{eqnarray}

\noindent
Using $\alpha\left(m^2_{\rho}\right) = 1$ and
$\alpha\left(m^2_{\rho_{3}}\right)=3$ and the median mass
values we find
     \begin{eqnarray}
     \alpha_0=0.479\;,\quad \alpha^{\prime}=0.881 (GeV)^{-2}
     \end{eqnarray}
     
\noindent
This then gives
     \begin{eqnarray}
     \alpha\left(m^2_{\pi}\right) =\alpha_0 +\alpha^{\prime} m^2_{\pi} = 0.497
     \end{eqnarray}

\noindent
which is remarkably close to Lovelace's quantization condition eqn.(4).

The s-wave scattering length $a^0_0$ and $a^2_0$ can be obtained
using definite isospin combinations of the amplitude eqn.(1) as
described in ref.[2] and [3].  We find [5]
     \begin{equation}
     a^0_0 = \displayfrac{\pi\beta\alpha^{\prime}m^2_{\pi}}{m_{\pi}}\cdot
     \frac{7}{2} \left\protect[1+\frac{20}{7}\ell n 2\left(\alpha^{\prime}
     m^2_{\pi}\right)
     +22.1\left(\alpha^{\prime}m^2_{\pi}\right)^2+\ldots\right\protect]
     \end{equation}

    \begin{equation}
     a^2_0=\displayfrac{-\pi\beta \alpha^{\prime} m^2_{\pi}}{m_{\pi}}
     \left\protect[ 1-4 \ell n 2\left(\alpha^{\prime} m^2_{\pi}\right)
     +(3.36) \left(\alpha^{\prime} m^2_{\pi}\right)^2
     +\ldots\right\protect]
     \end{equation}

The constant $\beta$ can be determined by relating the I=1
amplitue to the experimental width $\Gamma$ of the $\rho$ [3]
     \begin{eqnarray}
     \beta = \frac{3}{4} \displayfrac{\Gamma m^2_{\rho}}{q^{3}} \approx 1.45
     \end{eqnarray}

\noindent
Here $q$ is the C.M. momentum of the decaying pions.
Alternately instead of using the experimental $\rho$-width
one can use the famous KSRF relation\cite{7} to get
     \begin{subequations}
     \begin{eqnarray}
     \beta=\displayfrac{g^2_{\rho\pi\pi}}{8\pi} &=&\frac{1}{8\pi}\quad
     \displayfrac{2 m^2_{\rho}}{F^2_{\pi}}\quad\quad (F_{\pi}=184 MeV)\\
     &\simeq& 1.39
     \end{eqnarray}
     \end{subequations}

\newpage
\noindent
Multiplying eqn.(11a) by $\alpha^{\prime}$ and using
$2\alpha^{\prime}m_{\rho}^{2}=1$ we can write\footnote{Using
eqn.(12) in eqn.(8) and (9) makes the leading coefficients a
factor $\pi/2$ larger than the values given by Weinberg\cite{6}.
However if one keeps in eqn.(1) only the
$\alpha(s)=1\quad\sigma$-pole term one is lead back to
Weinberg's values.  See related comments in \cite{8}.  For a
recent discussion of Chiral perturbation theory for $\pi\pi$
scattering see\cite{9}.  The precise numerical value of $\beta$
plays no role in Lovelace quantization condition eqn.(4) and in
our discussion of PCAC in nuclear medium}.
     \begin{eqnarray}
     \beta \alpha^{\prime} =
     \frac{1}{8\pi}\quad\displayfrac{1}{F^2_{\pi}} 
     \end{eqnarray}

\noindent
which will be useful later.

Before we turn to the nuclear medium, it is useful to recall the
generalization of the Lovelace Quantization Condition eqn.(4) by
Ademollo, Veneziano and Weinberg\cite{8}.  They obtained
     \begin{eqnarray}
     \alpha_X(0) -\alpha_A(0) = 1/2\cdot
     \end{eqnarray}

\noindent
where the hadron $X$ couples to the hadron $A$ by pion with $X$
and $A$ having opposite normality.  (Normality = parity $\otimes
(-1)^J$ bosons; parity $\otimes (-1)^{J-1/2}$ for fermions).
Using eqn.(13) they deduced several relations
     \begin{eqnarray}
     \alpha^{\prime}\left(m^2_{K^{*}} - m^2_K\right) = 1/2\quad\quad
     \alpha^{\prime}\left(m^2_{\Delta} - m^2_N\right)=1/2
     \end{eqnarray}

\noindent
etc. besides of course the Lovelace quantization, condition.
     \begin{eqnarray*}
     \alpha^{\prime}\left(m^2_{\rho} -m^2_{\pi}\right)=1/2
     \end{eqnarray*}

All this predates QCD.  Let us now consider the question, how
$\alpha^{\prime}$ is determined in QCD.  Several years ago,
Nambu\cite{10} derived a string like equation for the path
ordered phase-factor
     \begin{eqnarray}
     U\protect[\sigma\protect] = p\exp \left(i\;\int\limits_{\sigma}
     A_{\mu} dz^{\mu}\right)
     \end{eqnarray}

\noindent
where $\sigma$ is a space-like curve and $A_{\mu}=g \sum A_{\mu}^a \lambda^a/2$.

By considering the variation in $U[\sigma]$ for a normal
displacement and iterating it Nambu obtained the equation
     \begin{eqnarray}
     \left(\displayfrac{\delta}{\delta\sigma_{\mu t}}
     \displayfrac{\delta}{\delta \sigma_{\mu t}} + C\right) U
     \protect[\sigma\protect] =0
     \end{eqnarray}

\noindent
with
     \begin{eqnarray}
     C = G_{\mu t} G_{\mu t}
     \end{eqnarray}

\noindent
where $\mu$ and $t$ refer to normal and tangential directions
along the string and
$G_{\alpha\beta}=\partial_{\alpha}A_{\beta}-\partial_{\beta}A_{\alpha}
-i[A_{\alpha},A_{\beta}]$
is the gluon field tensor, Identifying eqn.(16) with the string
equation from the Nambu-Goto Action leads to 
     \begin{eqnarray} 
     C =-\left(\displayfrac{1}{2\pi\alpha^{\prime}}\right)^2 
     \end{eqnarray}

\noindent
Nambu further pointed out that the energy density of the string
regarded as a chromoelectric flux tube is consistent in
magnitude with the value of the gluon condensate
$<0|G_{\alpha\beta} G^{\alpha\beta}|0>$ determined from QCD sum
rules\cite{11}.  Also he estimated the cross-section area of the
flux tube $a$ to be
     \begin{eqnarray}
     a \approx (0.5\;\mbox{fermi})^2
     \end{eqnarray}

The slope of the Regge trajectory $\alpha^{\prime}$ can also be
determined in the MIT bag model.  Johnson and Thorn\cite{12}
derived the relation
     \begin{eqnarray}
     \alpha^{\prime} =\displayfrac{1}{16\pi^{3/2}}\left(\frac{3}{2}
     \right)^{1/2}\displayfrac{1}{\sqrt{\alpha}_{s}\sqrt{B}}
     \end{eqnarray}

\noindent
which from the phenomenological values $\alpha_s=0.5$ and
$B^{1/4}=146 MeV$ used in Bag model studies yields
$\alpha^{\prime}=0.91(GeV)^{-2}$ in agreement with experiment.
The Bag  pressure $B$ can be readily interpreted as the
difference between the energy density of the physical vacuum
(outside of the Bag) and perturbative vacuum (inside of the Bag).
Now by the trace anomaly the energy density of the physical
vacuum  is related to the gluon condensate, $<0|G_{\mu\nu}G^{\mu\nu}|0>$.

Guided by the above considerations we shall assume
     \begin{eqnarray}
     \displayfrac{1}{(\alpha^{\prime})^{2}} = K < 0|G_{\mu\nu}G^{\mu\nu}|0>
     \end{eqnarray}

\noindent
where the precise value of the constant $K$ will not be
important in the following.

We now turn to the nuclear medium.  At low nuclear densities
chiral symmetry would remain broken even if its magnitude
changes significantly.  It follows from eqn.(19) that the
transverse string radius is small compared to internucleon
separation in a nucleus.  Also as emphasized by Migdal\cite{13}
the transition from weak coupling to strong coupling in $\alpha_s$ 
takes place at small distances 0.2 to 0.3 fm and the string picture
makes sense even for low angular momenta.  
It is then reasonable to extend the dual amplitude
to the nuclear case at least for low densities 
by replacing $\alpha(s)$ by
$\alpha_{\mbox{med}}(s)$ where we write
     \begin{eqnarray}
     \alpha_{\mbox{med}}(s) = \alpha_{0,\;\mbox{med}} + 
      \alpha^{\prime}_{\mbox{med}}\cdot s
      \end{eqnarray}

\noindent
By virtue of eqn.(21), the modified string tension
$\alpha^{\prime}_{\mbox{med}}$ can be obtained from
    \begin{eqnarray}
    \displayfrac{\left(\alpha^{\prime}_{\mbox{med}}\right)^2}
    {(\alpha^{\prime})^2} =
    \displayfrac{<0|G_{\alpha\beta}G^{\alpha\beta}|0>}
    {<0|G_{\alpha\beta} G^{\alpha\beta}|0>_{\mbox{med}}}
    = \lambda^2
    \end{eqnarray}

\noindent
where $<0|G_{\mu\nu}G^{\mu\nu}|0>_{\mbox{med}}$ denotes the
value of the gluon condensate in the nuclear medium.  
$\lambda$ is a convenient scale factor (in the medium $\lambda>1$ see
below).  PCAC or the Adler Zero in medium will now 
be satisfied by the modified Lovelace quantisation condition
     \begin{eqnarray}
     \alpha_{\mbox{med}}\left(m^2_{\pi}\right) = 1/2
     \end{eqnarray}

\noindent
Eqn.(23) and eqn.(24) are sufficient to derive a number of
results.  We indicate a few below.  We use an asterisk to denote
the in-medium value.  From eqn.(4), (14), (23) and (24) we have\cite{14}.
     \begin{eqnarray}
     \displayfrac{\left(m^{*2}_{\rho}-m^2_{\pi}\right)}
     {\left(m^{2}_{\rho}-m^{2}_{\pi}\right)} =
     \displayfrac{{\alpha}^{\prime}}{\alpha^{\prime}_{\mbox{med}}} =
     \frac{1}{\lambda}
     \end{eqnarray}

     \begin{eqnarray}
     \left(m^{* 2}_{\Delta} - m^{*2}_N\right)\big/
     \left(m^{2}_{\Delta}-m^2_N\right) =\frac{1}{\lambda}
     \end{eqnarray}

\noindent
The $\Delta-N$ mass difference must decrease within the medium.
Using eqn.(8), we find that the S-wave scattering length $a^0_0$ 
should increase inside the medium.
     \begin{eqnarray}
     a^{0 *}_0 = a^0_0\; \lambda
     \end{eqnarray}

>From eqn.(12) the pion-decay constant $F_{\pi}$ decreases inside the nucleus
     \begin{eqnarray}
     F^{*}_{\pi} = F_{\pi} \lambda^{-1/2}
     \end{eqnarray}

It is easy to understand Eqn.(25), (27) and (28).  In a picture
of the pion as a quark antiquark bound state, $F^*_{\pi}< F_{\pi}$
corresponds to decrease of the bound state wave function at the
origin or an increase in the radius.  Decrease of $|\psi(0)|^2$
in turn implies a reduction of the hyperfine splitting between the
singlet pion and the triplet $\rho$ and of course an
increase in the radius of the pion implies an
increase in the $\pi\pi$ cross section.

The scaling ratio $\lambda$ can be estimated as follows.  One
can write in the linear density approximation\cite{15}.
     \begin{eqnarray*}
     <0|G_{\mu\nu}G^{\mu\nu}|0>_{\mbox{med}} = 
     <0|G_{\mu\nu}G^{\mu\nu}|0> +\rho<N|G_{\mu\nu}G^{\mu\nu}|N>
     \end{eqnarray*}

\noindent
where $\rho$ is the density of nucleons inside the nucleus
($\rho\approx 0.16 fm^{-3}$).  The matrix element
$<N|G_{\mu\nu}G^{\mu\nu}|N>$ is related to nucleon mass by the
trace-anomaly.  One has\cite{16}
     \begin{eqnarray*}
     <N|G_{\mu\nu} G^{\mu\nu}|N> \approx -32 \pi^2 \otimes  78 MeV
     \end{eqnarray*}

\noindent
Using the QCD sum rule estimate\cite{11}
 $<0|G_{\mu\nu}G^{\mu\nu}|0> \approx 0.5(GeV)^4$
this means $\lambda \approx (1.06)^{1/2}$ at ordinary nuclear
densities.  In a heavy ion collision we can expect $\lambda$ to
be significantly larger than 1.

One can also look for other experimental signatures.  For
example, the transverse momentum distribution of secondaries in
high energy collisions is exponentially supressed.  This
damping factor is proportional to 
$(\alpha^{\prime})^{-1/2}$.  With increasing
$\alpha^{\prime}_{\mbox{med}}$, we should expect less damping i.e., 
the average transverse momentum of the secondaries  increases with density
or equivalently centre mass energy of the collision.


\begin{thebibliography}{99}
\bibitem{5} These are easily obtained from the  expansions 
for $1/\Gamma(x)$ as $x\rightarrow 0$ and $\Gamma(1/2+y)$ as $y\rightarrow 0$.
\end{thebibliography}
\end{document}